\documentclass[aps,amsmath,amssymb,twocolumn,longbibliography,10pt,prx,superscriptaddress]{revtex4-2}
\usepackage{graphicx}
\usepackage[colorlinks=true,linkcolor=blue,citecolor=blue,urlcolor=blue]{hyperref}
\usepackage{color}
\usepackage{multirow}
\usepackage{soul}
\usepackage{tabularx}
\newcommand{\la}{\langle}

\newcommand{\ua}{\uparrow}
\newcommand{\da}{\downarrow}
\newcommand{\Ua}{\Uparrow}
\newcommand{\Da}{\Downarrow}
\newcommand{\micron}{\mathrm{\mu m}}
\newcommand{\CC}{\mathrm{C}}
\newcommand{\EE}{\mathrm{E}}
\newcommand{\En}{\mathcal{E}}

\newcommand{\ket}[1]{| {#1} \rangle}
\newcommand{\bra}[1]{\langle {#1} |}
\newcommand{\Edc}{E_\mathrm{dc}}
\newcommand{\vecEdc}{\vec{E}_\mathrm{dc}}

\usepackage{times}

\begin{document}

\title{Tunable two-species spin models with Rydberg atoms in circular and elliptical states}
\author{Jacek Dobrzyniecki}
\email{Jacek.Dobrzyniecki@oist.jp}
\affiliation{Faculty of Physics, University of Warsaw, Pasteura 5, 02-093 Warsaw, Poland}
\affiliation{Quantum Systems Unit, Okinawa Institute of Science and Technology Graduate University, Okinawa 904-0495, Japan}
\author{Paula Heim}
\affiliation{Faculty of Physics, University of Warsaw, Pasteura 5, 02-093 Warsaw, Poland}
\affiliation{Mathematical Institute, University of Oxford, Oxford, OX2 6GG, UK}
\author{Michał Tomza}
\email{Michal.Tomza@fuw.edu.pl}
\affiliation{Faculty of Physics, University of Warsaw, Pasteura 5, 02-093 Warsaw, Poland}
\date{\today}

\begin{abstract}
We propose a scheme for constructing versatile quantum simulators using ultracold Rydberg atoms in long-lived circular and elliptical states. By exciting different subspaces of internal atomic states, the atoms can be used to simulate two effective spin species with different spin-spin interactions. The strengths of transverse and longitudinal spin-spin interactions, both intra- and inter-species, can be controlled within a wide range of values. This setup can be used to simulate two-species spin models or lattice models with two sublattices. We show examples of specific models which can be realized.
\end{abstract}

\maketitle


\section{Introduction}\label{sec:Intro}

Quantum simulation allows to investigate complex physical models, with an unprecedented level of control and insight into the system properties. Technological advancements now allow for the construction of quantum simulators using arrays of ultracold particles trapped in precisely positioned optical tweezers~\cite{2018-Barredo-Nature,2019-Anderegg-Science,2021-Altman-PRXQ,2022-Jenkins-PRX,2022-Bluvstein-Nature,2022-Zhang-QSciTech,2024-Gyger-PRR}. The study of spin models is a crucial application of these simulators. Spin model simulators can be realized by treating internal levels of the particles as effective ``spin'' states, with interactions between particles mapped onto effective ``spin interactions''. Previous proposals for spin model simulators include setups made up of ultracold polar molecules~\cite{2024-Cornish-NatPhys}, atoms~\cite{2017-Gross-Science}, or their mixtures~\cite{2023-Dobrzyniecki-PRA}.

Atoms in highly-excited Rydberg states are especially promising for quantum simulations due to their strong, tunable interactions, including complex anisotropic dipolar couplings~\cite{2010-Saffman-RevModPhys,2019-Adams-JPB}. Various studies have explored the use of Rydberg atom systems for quantum information processing~\cite{2000-Jaksch-PRL,2001-Lukin-PRL,2010-Isenhower-PRL}, simulation of spin models~\cite{2010-Weimer-NatPhys,2013-Gunter-Science,2015-Barredo-PRL,2018-Schauss-QST,2018-DeLeseleuc-PRL,2020-Browaeys-NatPhys,2020-Celi-PRX,2021-Wu-CPB,2024-JuliaFarre-PRA-Quantized,2024-Mukherjee-Arxiv}, and simulation of hard-core bosons in a lattice~\cite{2024-Poon-AdvQuantumTech}. The simulators based on Rydberg atoms have possible uses for studying quantum magnetism~\cite{2024-JuliaFarre-PRA-Amorphous,2024-Homeier-Arxiv,2024-Liu-Arxiv-Supersolidity} and topological phases~\cite{2019-DeLeseleuc-Science,2023-Kornjaca-CommPhys,2024-Zeng-Arxiv}. Notably, a recent paper considers a Rydberg atom array composed of two different atom species, which can exhibit more varied interactions than single-species arrays~\cite{2024-Liu-Arxiv-Novel}. In addition to theoretical proposals, Rydberg atom-based quantum simulators have seen promising experimental realizations~\cite{2015-Barredo-PRL,2016-Labuhn-Nature,2017-Bernien-Nature,2019-DeLeseleuc-Science,2020-Schlosser-JPB,2021-Semeghini-Science,2021-Scholl-Nature,2021-Ebadi-Nature,2021-Signoles-PRX,2022-Scholl-PRXQ,2023-Steinert-PRL,2024-Bornet-PRL}.

Among the various Rydberg states, \emph{circular} states have attracted considerable attention. Circular states have the maximum possible magnetic quantum number ($|m_\ell| = \ell = n-1$, where $\ell$, $n$ are the azimuthal and principal quantum numbers). They exhibit notably long lifetimes due to limited radiative decay pathways. The radiative lifetimes of circular states scale as $n^5$, and typical circular level lifetimes are on the order of $10^{-2}\,\mathrm{s}$ (\emph{e.g.} $2.5\times10^{-2}\,\mathrm{s}$ for $n = 48$)~\cite{2018-Nguyen-PRX}. In comparison, the lifetimes of low-$\ell$ Rydberg states scale approximately as $n^3$; for $n = 48$ they are on the order of $10^{-4}\,\mathrm{s}$~\cite{1994-Gallagher-Book}. The lifetime of circular states can be further extended by placing the atoms inside small cavities that do not support the modes corresponding to their decay wavelengths~\cite{1981-Kleppner-PRL,1985-Hulet-PRL}. Techniques for preparing circular Rydberg states are well-established~\cite{1983-Hulet-PRL,1986-Liang-PRA,2018-Morgan-PRA,2020-CantatMoltrecht-PRR}, with recent advancements allowing for precise positioning of these atoms in arrays~\cite{2023-Ravon-PRL,2024-Holzl-PRX}. Very recently, dipolar interactions between trapped circular atoms have been observed experimentally~\cite{2024-Mehaignerie-Arxiv}. These facts make circular Rydberg states a promising tool for quantum simulation~\cite{2023-Mehaignerie-Thesis}, and the rapid advancement of experimental techniques makes circular states a timely subject.

Several proposals have explored practical applications of circular Rydberg states. For instance, some proposals have suggested using the entire ladder of Rydberg levels, from low-$\ell$ to circular levels, to simulate the states of an arbitrarily large spin~\cite{2014-Signoles-NatPhys,2022-Kruckenhauser-QST}. Recently, detailed proposals have been published for using systems of cold circular Rydberg atoms in quantum simulation~\cite{2018-Nguyen-PRX} and computation~\cite{2021-Cohen-PRXQ}. These proposals use a scheme where the two spin-1/2 states are encoded in two circular states. This approach typically yields effective spins with strong spin-exchanging couplings, in contrast with weaker longitudinal couplings~\cite{2023-Mehaignerie-Thesis}.

In this paper, we investigate a different scheme, where the two spin-1/2 states are encoded as a circular level and an elliptical (\emph{i.e.} $|m_\ell| = n-2$) level. This results in effective spins for which longitudinal spin-spin couplings are much stronger than spin-exchanging couplings. We evaluate and compare the interaction strengths in both encoding schemes. Furthermore, we explore the possibility of combining both schemes to simulate a ``two-species'' spin model with tunable inter- and intra-species interactions. Such a model can be simulated with an array of identical atoms, flexibly assigned to either ``spin species'' as desired. This simplifies preparation and allows the same atom arrangement to simulate systems with various species populations. We furthermore demonstrate two example systems which can be realized in this approach.

We note that elliptical Rydberg states have received less attention in the literature, being treated mostly as sources of unwanted couplings of circular states. Our work finds a practical application for elliptical states, providing an initial exploration of their potential for new quantum simulation schemes.

This paper is organized as follows. In Section~\ref{sec:model}, we describe a physical system of ultracold atoms and outline the two distinct mapping schemes for creating effective spins. In Section~\ref{sec:effective-interactions}, we analyze the resulting effective spin interactions in these two distinct spin species. In Section~\ref{sec:example-systems}, we showcase example models which can be created with this approach. In Section~\ref{sec:experimental}, we discuss the experimental feasibility of realizing the proposed two-species setup. Section~\ref{sec:conclusion} is the conclusion.


\section{The Model}\label{sec:model}

\subsection{The physical atomic system}

We consider a system of $N$ atoms, modeled as point-like particles and numbered as $j=1,\ldots,N$. The atoms are placed at fixed locations $\vec{R}_j$. The Hamiltonian describing this system is
\begin{equation}
\label{eq:physical_hamiltonian}
    \hat{H} = \sum\limits_{j=1}^N \hat{h}^{(j)} + \sum\limits_{k=1, j<k}^N \hat{V}^{(j,k)}, 
\end{equation}
where $\hat{h}^{(j)}$ is the one-body Hamiltonian describing the internal levels of atom $j = 1,...,N$, and $\hat{V}^{(j,k)}$ describes the interaction between atoms $j$ and $k$. We assume all $N$ atoms are described by identical one-body Hamiltonians $\hat{h}^{(j)}$. Additionally, we assume the system is made up of alkali-metal atoms, which are a common choice in cold-atom experiments and quantum simulators due to their favorable properties. The exact choice of atomic element is inconsequential, since circular-state valence electrons feel very little interaction with the nucleus, and their properties depend only negligibly on the atomic number.

The system is subjected to a uniform electric field $\vecEdc$ and a magnetic field $\vec{B}$, both aligned along the $Z$-axis (\emph{i.e.}, the quantization axis), of strengths $\Edc$ and $B$ respectively. The effects of $\vecEdc$ and $\vec{B}$ on the atomic levels are incorporated in $\hat{h}^{(j)}$.

Throughout the paper, we assume the magnetic field strengths are constrained to below 1000 Gauss, as higher values are challenging to obtain in ultracold-atoms laboratory settings. Similarly, the maximum electric field strength is limited in practice, because at high $\Edc$ Rydberg atoms undergo spontaneous ionization. Note that circular states can withstand higher electric fields compared with low-$\ell$ states of the same $n$. As an example, in Ref.~\cite{1983-Hulet-PRL}, the ionization thresholds for circular $n=19$ Li states were estimated as $6\,\mathrm{kV/cm}$, compared with $4.6\,\mathrm{kV/cm}$ for $m_\ell = 2$ states of same $n$. Similarly, Ref.~\cite{2020-Teixeira-PRL} gives the threshold field as $\approx 125\,\mathrm{V/cm}$ for the $n=51$ circular level of Sr, compared with $\approx 50\,\mathrm{V/cm}$ for low-$\ell$ states. In our paper, we focus on levels up to $n = 73$ and, extrapolating from the data above, we limit the considered electric fields to $20\,\mathrm{V/cm}$.

\subsection{Internal atomic levels}

We use the $\ket{n,\ell,m_\ell}$ basis to describe the internal atomic levels, where $n$ is the principal quantum number of the valence electron, $\ell$ is its angular orbital momentum, and $m_\ell$ is the projection of $\ell$ on the quantization axis. As the spin-orbit splitting is insignificant for circular states, we can ignore the fine structure. Circular states are those which have the maximum possible $|m_\ell| = \ell = n-1$. For alkali-metal atoms, in a given $n$ manifold one can define two circular states with $m_\ell = \pm \ell$. We are also interested in the \emph{elliptical} states, characterized by $|m_\ell| = n-2$. One can define four elliptical states for given $n$ [with $\ell = (n-2)$ or $\ell = (n-1)$, and $m_\ell = \pm \ell$]. 

In the zero-field limit, eigenstates of $\hat{h}^{(j)}$ can be written as $\ket{n,\ell,m_\ell}$, but states with different $m_\ell$ are degenerate and thus $m_\ell$ is ill-defined as a quantum number. For higher fields, this degeneracy is lifted and $m_\ell$ becomes a good quantum number, even though the eigenstates become superpositions of levels with different $n$ and $\ell$.
     
Throughout this paper, we define $\ket{n\CC_\pm}$ as the field-dressed eigenstate of $\hat{h}^{(j)}$ which approaches the form ${|n, \ell=(n-1), m_\ell = \pm (n-1)\rangle}$ in the small-field limit. Similarly, we define $\ket{n\EE_\pm}$ as the field-dressed eigenstate of $\hat{h}^{(j)}$ which approaches the form ${|n, \ell=(n-2), m_\ell = \pm (n-2)\rangle}$ in the small-field limit. [For sake of simplicity, we do not consider elliptical states with $\ell = (n-1)$.]
     
For given field strengths $\Edc$, $B$, we find the compositions and energies of field-dressed eigenstates by constructing and numerically diagonalizing the one-body Hamiltonian, using procedures from the Python library {\tt pairinteraction}~\cite{2017-Weber-JPB}.

\subsection{The interaction term}

For simplicity, in the two-body interaction terms $\hat{V}^{(j,k)}$ we only consider electric dipole-dipole interactions. We omit the higher-order multipole-multipole interactions, which are generally weaker than dipole interactions by one or more orders of magnitude~\cite{2021-Cohen-PRXQ}. We denote the electric dipole operator acting on atom $j$ as $\hat{d}^{(j)}$. Its components $\hat{d}^{(j)}_q$ can be written in terms of the spherical coordinate system, where basis vectors are labeled by $q = 0,\pm1$: $\vec{e}_0 \equiv \vec{e}_Z, \vec{e}_{\pm 1} \equiv \mp (\vec{e}_X \pm i \vec{e}_Y) / \sqrt{2}$. The dipole-dipole interaction term can then be written succinctly as follows~\cite{2003-Brown-Book}:
\begin{equation}
 \label{eq:dipole-dipole-short-form}
 \hat{V}^{(j,k)} = \sum_{q',q=-1}^{+1} v^{(j,k)}_{q';q} \hat{d}^{(j)}_{q'} \hat{d}^{(k)}_q,
 \end{equation}
 where
 \begin{align}
  v^{(j,k)}_{0;0} = 2 v^{(j,k)}_{+1;-1} = 2 v^{(j,k)}_{-1;+1} &= \frac{1-3 \cos^2 \theta_{jk}}{4 \pi \epsilon_0 |R_{jk}|^3}, \label{eq:dipole-dipole-angular-factors} \\
  v^{(j,k)}_{0;\pm 1} = v^{(j,k)}_{\pm1;0} &= \pm \frac{3}{\sqrt{2}} \frac{ \sin\theta_{jk}\cos\theta_{jk} } {4 \pi \epsilon_0 |R_{jk}|^3} e^{\mp i\phi_{jk}}, \nonumber \\
  v^{(j,k)}_{\pm 1; \pm 1} &= -\frac{3}{2} \frac{  \sin^2\theta_{jk} } {4 \pi \epsilon_0 |R_{jk}|^3} e^{\mp i2\phi_{jk}}. \nonumber 
 \end{align}
Here, $\vec{R}_{jk} = \vec{R}_k - \vec{R}_j$, \emph{i.e.} the distance between atoms $j$ and $k$. $\theta_{jk}$ is the polar angle of $\vec{R}_{jk}$ ($\cos \theta_{jk} = \vec{R}_{jk}\cdot \vec{e}_0/|R_{jk}|$), and $\phi_{jk}$ is the azimuthal angle of rotation about the $\vec{e}_0$ axis (relative to an arbitrarily chosen $X$ axis). 

Per the electric dipole selection rules (${ \Delta \ell = \pm 1, \Delta m_\ell = 0,\pm 1 }$), the dipole operators allow the coupling between neighboring circular levels ${ \ket{n\CC_\pm} \leftrightarrow \ket{(n+1)\CC_\pm} }$. The corresponding transition dipole moments scale as ${ \sim n^2 e a_0 }$~\cite{1994-Gallagher-Book}. Similarly, the couplings ${ \ket{n\CC_\pm} \leftrightarrow \ket{n\EE_\pm} }$ and ${ \ket{n\CC_\pm} \leftrightarrow \ket{(n+2)\EE_\pm} }$ are dipole-allowed, with a matrix element scaling as ${ \sim n^{3/2} e a_0 }$~\cite{1994-Gallagher-Book}. The coupling ${ \ket{n\CC_\pm} \leftrightarrow \ket{(n+1)\EE_\pm} }$ becomes dipole-allowed in an electric field, because the field-dressed level ${ \ket{(n+1)\EE_\pm} }$ contains a large admixture of an elliptical state with ${ \ell = n }$; the resulting matrix element likewise scales as ${ \sim n^{3/2} e a_0 }$. 

\subsection{Choosing the effective spin basis}

In quantum simulation of $S$-spin particles, a common strategy is to encode the $2S+1$ spin states in a particular subspace of atomic levels. We expand upon this approach by defining two separate subspaces, which correspond to two distinct spin species. If we select two subspaces which significantly differ with regards to transition matrix elements between the constituent levels, the atoms can effectively represent two distinct spin species with characteristic inter- and intra-species interaction strengths. This scheme allows to map identical atoms to non-identical spin species, simply by preparing them in the appropriate subspaces. In the following discussion, we focus on simulating 1/2-spin particles, so that each chosen subspace will consist of two levels. Extending this scheme to higher $S$ may be possible, although likely presents additional complications. 

For the first spin species, we choose a pair of neighboring circular states $\ket{n\CC_\pm}$ and $\ket{(n+1)\CC_\pm}$. These states can be mapped, in arbitrary order, to the 1/2-spin states $\ket{\Da}$ and $\ket{\Ua}$. We label their energies as $\En_\Da$ and $\En_\Ua$. This assignment of states represents the established approach to effective spin simulation, known from previous literature~\cite{2018-Nguyen-PRX,2021-Cohen-PRXQ}. 

For the second spin species, we choose a different circular state $\ket{n' \CC_\pm}$, paired with either $\ket{(n'+1) \EE_\pm}$ or $\ket{(n'+2) \EE_\pm}$. The circular and elliptical state can be mapped, also in arbitrary order, to the 1/2-spin states $\ket{\da}$ and $\ket{\ua}$, with energies $\En_\da$ and $\En_\ua$.

We refer to these two types of effective spins as the CC species and the CE species, respectively. For brevity, we use the term ``CC atom'' to designate an atom excited into the first subspace, and similarly for ``CE atom''.

Spin-exchanging interactions between different species are often a desirable part of the system dynamics. The spin exchange between CC and CE species depends on the F\"orster defect ${\Delta = (\En_\Ua - \En_\Da) - (\En_\ua - \En_\da)}$, which expresses the energy cost of a single exchange. To enable resonant spin exchange, $\Delta$ should be comparable to, or smaller than, the spin exchange matrix element. This can be achieved by tuning the external electric and magnetic fields to minimize $\Delta$. Throughout the rest of this paper, whenever analyzing the system properties in different fields, we consistently adjust $\Edc$ and $B$ together in order to minimize $\Delta$. For given $\Edc$, we define $B_\mathrm{res}$ as the value of $B$ where the F\"orster defect is $\Delta = 0$. In practice, since realistically the ability to control $B$ is limited by environmental field fluctuations and other sources of error, we define $B_\mathrm{res}$ only with an accuracy down to $0.01\,\mathrm{Gauss}$, so $\Delta$ in our calculations has a residual nonzero value.

Achieving interspecies spin exchange places constrains on the choice of CC and CE effective spin levels. The values of $n$ and $n'$ should be chosen in such a way that the transition frequencies $\En_\Ua-\En_\Da$ and $\En_\ua-\En_\da$ are similar in zero field, and therefore can be tuned into resonance via small external fields (on the order of $\sim 1\,\mathrm{V/cm}$ and $\sim 100\,\mathrm{Gauss}$). For the scheme $n \rightarrow n+1, n' \rightarrow n'+1$, the values of $n$ and $n'$ should therefore be similar, while for the scheme $n \rightarrow n+1, n' \rightarrow n'+2$, generally one has $n' \gg n$. 

Additionally, interspecies spin exchange constrains the assignment of the spin level labels: if for species CC we assign the higher-energy state to $\Ua$ and the lower-energy state to $\Da$, then for species CE we need to assign $\ua, \da$ in the same way, so that the pair states $\ket{\Ua \da}$ and $\ket{\Da \ua}$ can have similar total energy. The sign of $m_\ell$ picked for the two species can be the same or different. However, picking level pairs with opposite signs of $m_\ell$ (\emph{e.g.}, ${ \ket{n \CC_- },\ket{(n+1) \CC_-} }$ and ${ \ket{n' \CC_+ }, \ket{(n'+2) \EE_+ } }$) helps ensure that the Zeeman shifts of the transition frequencies have different signs, which makes it easier to match the transition frequencies. Additionally, the extreme difference in $m_\ell$ helps to prevent mixing between the two spin state subspaces.

It is worth noting that the electric field $\vecEdc$, acting along the $Z$-axis, has a different effect on circular and elliptical states~\cite{1994-Gallagher-Book}. Specifically, circular states lack a first-order Stark shift, which is normally caused by mixing to nearby levels with the same $m_\ell$ and opposite symmetry; a circular state $\ket{n \CC}$ has no partners with the same $m_\ell$ within the same $n$ manifold. For this reason, in the relatively small fields $\Edc \sim 1\,\mathrm{V/cm}$ that we consider, the energy and composition of $\ket{n \CC_\pm}$ are only negligibly affected by $\Edc$ (only a weak second-order Stark effect occurs). On the other hand, the elliptical states $\ket{n \EE_\pm}$ display a much stronger first-order Stark shift. This will be shown to be particularly relevant for the effective spin interactions and their tunability via electric fields.

Finally, we emphasize that a two-species model with inter-species spin exchange cannot be easily realized using only circular states, \emph{i.e.}, using the subspaces $\ket{n \CC} \rightarrow \ket{(n+1) \CC}$ and $\ket{n' \CC} \rightarrow \ket{(n'+1) \CC}$. In this approach, to make the two species meaningfully different, it is necessary to use highly different $n, n'$ (to obtain different interaction strengths). However, this is precluded by the requirement that transition frequencies $\En_\Ua-\En_\Da$ and $\En_\ua-\En_\da$ are similar in zero field.

\section{Example interaction parameters}\label{sec:effective-interactions}

\subsection{Obtaining the effective spin Hamiltonian}

Our goal is to recast the physical Hamiltonian of Eq.~\eqref{eq:physical_hamiltonian} into an effective spin~Hamiltonian. The most complete form of this effective Hamiltonian, incorporating all the possible one- and two-body spin terms, is
\begin{align}
\label{eq:spin_hamiltonian}
 \hat{H}_\mathrm{eff} &= \left(\sum_{j=1}^N C^{j}_I\right) \hat{\mathbb{I}} \\
 &+ \sum_{j=1}^N C^{j}_z \hat{S}^{(j)}_z \nonumber \\
 &+ \sum_{k=1, j<k}^N \left[C^{jk}_{+-} \hat{S}^{(j)}_+ \hat{S}^{(k)}_- + (C^{jk}_{+-})^* \hat{S}^{(j)}_- \hat{S}^{(k)}_+ \right] \nonumber \\
 &+ \sum_{k=1, j<k}^N \left[C^{jk}_{++} \hat{S}^{(j)}_+ \hat{S}^{(k)}_+ + (C^{jk}_{++})^* \hat{S}^{(j)}_- \hat{S}^{(k)}_- \right] \nonumber \\
 &+ \sum_{j=1}^N \left[C^{j}_{+} \hat{S}^{(j)}_+ + (C^{j}_{+})^* \hat{S}^{(j)}_- \right] \nonumber \\
 &+ \sum_{j\ne k=1}^N \left[C^{jk}_{+z} \hat{S}^{(j)}_+ \hat{S}^{(k)}_z + (C^{jk}_{+z})^* \hat{S}^{(j)}_- \hat{S}^{(k)}_z \right] \nonumber \\
 &+ \sum_{k=1, j<k}^N C^{jk}_{zz} \hat{S}^{(j)}_z \hat{S}^{(k)}_z. \nonumber 
\end{align}
The most important terms include: effective Zeeman field couplings with coefficients $C^{j}_z$, transverse (spin-exchange) interactions with strengths $C^{jk}_{+-}$, and longitudinal interactions with strengths $C^{jk}_{zz}$. All the interaction coefficients $C$ have units of energy (we express the energy in units of $h \times \mathrm{Hz}$, with Planck's constant $h$ set to unity). The non-spin-conserving terms with coefficients $C^j_+$, $C^{jk}_{+z}$, and $C^{jk}_{++}$ have been included for completeness, but they typically have negligible impact on dynamics because they correspond to highly off-resonant processes. The constant term $\sum_j C^{j}_I$ merely shifts the energy reference and will be omitted in subsequent discussions.

The coefficients in Eq.~\eqref{eq:spin_hamiltonian} are obtained by evaluating the two-body interaction Hamiltonian
\[
\hat{H}^{(j,k)}_2 = \hat{h}^{(j)} + \hat{h}^{(k)} + \hat{V}^{(j,k)}
\]
for each pair of atoms $(j,k)$, restricting it to the subspace of effective spin states, and extracting the effective interaction coefficients from its matrix elements.

The procedure for each $(j,k)$ is as follows. We build an extensive basis of pair states, $\mathcal{B}_2^{(j,k)} = \{ \ket{\psi}^{(j)} \otimes \ket{\psi'}^{(k)} \}$, where each state is a tensor product of one-body eigenstates of $\hat{h}^{(j)}$ and $\hat{h}^{(k)}$. This basis is partitioned into two subspaces: the spin subspace $\mathcal{P}_2^{(j,k)}$, comprising the four states
\[
\mathcal{P}_2^{(j,k)} = \{ \ket{\ua^j \ua^k}, \ket{\ua^j \da^k}, \ket{\da^j \ua^k}, \ket{\da^j \da^k} \},
\]
or their equivalents for different species combinations; and the complementary subspace $\mathcal{Q}_2^{(j,k)}$, containing all other possible pair states. For computational feasibility, we limit $\mathcal{Q}_2^{(j,k)}$ to a selected subset, typically approximately $10^4$ states with energies and quantum numbers close to those in $\mathcal{P}_2^{(j,k)}$. We express the full two-body Hamiltonian,
\[
   \hat{H}_2^{(j,k)} = \hat{h}^{(j)} + \hat{h}^{(k)} + \hat{V}^{(j,k)},
\]
as a matrix in the basis $\mathcal{B}_2^{(j,k)}$, calculating all matrix elements directly using the {\tt pairinteraction} package. We then apply the Schrieffer-Wolff transformation to obtain the effective Hamiltonian $\hat{H}^{(j,k)}_{2,\mathrm{eff}}$, which is confined to the subspace $\mathcal{P}^{(j,k)}_2$ but incorporates the perturbative effects arising from the ${\mathcal{P}^{(j,k)}_2 \leftrightarrow \mathcal{Q}^{(j,k)}_2}$ couplings. The Schrieffer-Wolff transformation is performed using the procedure described in Ref.~\cite{2011-Bravyi-AnnPhys}. 

The Schrieffer-Wolff transformation is equivalent to perturbatively expanding an effective Hamiltonian, by treating the diagonal blocks ${\mathcal{P}^{(j,k)}_2 \leftrightarrow \mathcal{P}^{(j,k)}_2}$ and ${\mathcal{Q}^{(j,k)}_2 \leftrightarrow \mathcal{Q}^{(j,k)}_2}$ of $\hat{H}^{(j,k)}_2$ as the unperturbed Hamiltonian, and the off-diagonal block ${\mathcal{P}^{(j,k)}_2 \leftrightarrow \mathcal{Q}^{(j,k)}_2}$ as the perturbation. Therefore, for ${\ket{p},\ket{p'} \in \mathcal{P}^{(j,k)}_2}$, the matrix elements of $\hat{H}^{(j,k)}_{2,\mathrm{eff}}$ are given by~\cite{2011-Bravyi-AnnPhys}
\begin{align}
\label{eq:H2Eff_as_series}
\bra{p} \hat{H}^{(j,k)}_{2,\mathrm{eff}} \ket{p'} &= \delta_{p,p'} E_p + \bra{p} V^{(j,k)} \ket{p'} \\
&+ \sum_{\ket{q} \in \mathcal{Q}^{(j,k)}_2} \left[ \frac{\bra{p} V^{(j,k)} \ket{q} \bra{q} V^{(j,k)} \ket{p'} }{2} \right. \nonumber\\
&\times \left. \left( \frac{1}{E_p - E_q} + \frac{1}{E_{p'} - E_q} \right) \right] \nonumber\\
&+ O([V^{(j,k)}]^3), \nonumber
\end{align}
where the sum over $\ket{q}$ includes all the pair states $\ket{q} \in \mathcal{Q}^{(j,k)}_2$. ${ E_q = \bra{q} ( \hat{h}^{(j)} + \hat{h}^{(k)} ) \ket{q} }$ is the sum of single-particle energies of the states comprising $\ket{q}$. Note that $\bra{p} V^{(j,k)} \ket{p'}$ is the first-order dipolar contribution which falls off with distance as $1/R^3$, while the second-order contribution constitutes the van~der~Waals interaction that falls off as $1/R^6$. 

The obtained effective Hamiltonian [Eq.~\eqref{eq:H2Eff_as_series}] can be separated into a one-body component and an effective two-body interaction, $\hat{V}^{(j,k)}_{\mathrm{eff}}$:
     $$
     \bra{p} \hat{H}^{(j,k)}_{2,\mathrm{eff}} \ket{p'} = \delta_{p,p'} E_p + \bra{p} \hat{V}^{(j,k)}_{\mathrm{eff}} \ket{p'},
     $$
and the total effective spin~Hamiltonian in Eq.~\eqref{eq:spin_hamiltonian} can be expressed as
     \begin{equation}
     \label{eq:total_effective_spin_H}
         \hat{H}_\mathrm{eff} = \sum_{j=1}^N \hat{h}^{(j)} + \sum_{j < k} \hat{V}^{(j,k)}_\mathrm{eff}.
     \end{equation}

All the effective spin interaction coefficients can be therefore calculated from the appropriate interaction matrix elements:
\begin{align}
\label{eq:coefficients}
     C^{j}_z &= \left\langle \ua^j \left| \hat{h}^{(j)} \right| \ua^j \right\rangle - \left\langle \da^j \left| \hat{h}^{(j)} \right| \da^j \right\rangle \nonumber \\
     &+ \sum_{k \ne j}^N \frac{ U^{jk}_{\ua\ua} + U^{jk}_{\ua\da} - U^{jk}_{\da\ua} - U^{jk}_{\da\da} } {2}, \\
  C^{j}_{+} &= \sum_{k \ne j}^N \frac{\left\langle \ua^j\ua^k \left| \hat{V}^{(j,k)}_\mathrm{eff} \right| \da^j\ua^k \right\rangle + \left\langle \ua^j\da^k \left| \hat{V}^{(j,k)}_\mathrm{eff} \right| \da^j\da^k \right\rangle}{2}, \\
    C^{jk}_{+-} &= \left\langle \ua^j \da^k \left|  \hat{V}^{(j,k)}_\mathrm{eff} \right| \da^j \ua^k \right\rangle, \\
    C^{jk}_{++} &= \left\langle \ua^j \ua^k \left|  \hat{V}^{(j,k)}_\mathrm{eff} \right| \da^j \da^k \right\rangle, \\
  C^{jk}_{+z} &= \left\langle \ua^j\ua^k \left| \hat{V}^{(j,k)}_\mathrm{eff} \right| \da^j\ua^k\right\rangle - \left\langle \ua^j\da^k \left| \hat{V}^{(j,k)}_\mathrm{eff} \right| \da^j\da^k\right\rangle, \\
  C^{jk}_{zz} &= U^{jk}_{\ua\ua} - U^{jk}_{\ua\da} - U^{jk}_{\da\ua} + U^{jk}_{\da\da},
\end{align}
where $\ket{\ua^j}, \ket{\da^j}$ stand for the CC spin states $\ket{\Ua}, \ket{\Da}$ or the CE spin states $\ket{\ua}, \ket{\da}$, depending on the spin species assigned to atom $j$. The coefficients $U$ are the diagonal interaction elements $U^{jk}_{\sigma,\sigma'} = {\left\langle \sigma^j \sigma'^k \left|  \hat{V}^{(j,k)}_\mathrm{eff} \right| \sigma^j \sigma'^k \right\rangle}$.

Accurate spin system simulation requires that the spin subspace $\mathcal{P}^{jk}_2$ remains effectively decoupled from the residual subspace $\mathcal{Q}^{jk}_2$. Otherwise, the spin subspace is not closed and the atoms will not behave as two-level systems, causing simulation errors. In our calculations, we verify this condition in the following way: we diagonalize the full two-body Hamiltonian $\hat{H}_2^{(j,k)}$, and verify the overlap of each eigenstate with the spin subspace $\mathcal{P}_2^{(j,k)}$. For each eigenstate $\ket{\Phi}$, this overlap is defined as $\sum_{\ket{p} \in  \mathcal{P}_2^{(j,k)}} |\langle p \ket{\Phi}|^2$. We define a measure $\kappa^{(j,k)}$ as the fourth-largest overlap value among all eigenstates. If $\kappa^{(j,k)} \approx 1$ for all $j,k$, the requirement is satisfied.

Although $\kappa^{(j,k)}$ does not directly quantify simulation fidelity over time, it serves as a quantitative indicator which allows comparing different parameter sets from that perspective. Roughly, $1-\kappa^{(j,k)}$ can be interpreted as the probability of a single spin~pair experiencing leakage during one characteristic period of evolution. For an ensemble of $N \sim 10$ atoms, heuristic estimates suggest that maintaining all $\kappa^{(j,k)} \gtrsim 0.99$ is necessary to keep the probability of one or more error per period below $10\%$. For larger systems, better values of $\kappa$ are necessary.

Due to the complexity of the Rydberg atom spectrum, predicting $\kappa^{(j,k)}$ in advance is difficult and its value can greatly fluctuate as the experimental parameters are adjusted. In particular, at certain values of $\Edc$, $\kappa^{(j,k)}$ may be lowered by field-induced resonances between spin and non-spin pair states. We find that the most significant contribution to spin subspace contamination comes from the doubly-elliptical pair~state $\ket{n'' \EE_\pm}\ket{n'' \EE_\pm}$ mixing with non-spin states, particularly due to its resonances with certain states of the form~$\ket{n'' \CC_\pm}\ket{n''\mathrm{EE}_\pm}$ (where $\ket{n''\mathrm{EE}}$ designates a state with $|m_\ell| = n''-3$). In general, $\kappa^{(j,k)}$ can be enhanced by increasing inter-atomic distances and decreasing $n$, in order to weaken the couplings and increase the level spacing.

\subsection{Example values of effective spin coefficients} \label{sec:typical-values}

In the following section, we analyze example values of the spin interaction coefficients $C_{+-}, C_{zz}$, as well as the effective magnetic fields $C_z$ acting on individual spins. We focus on their dependence on the inter-atomic angle $\theta$ and external field $\Edc$. We exclude the coefficients $C_+$, $C_{++}$, and $C_{+z}$ from this analysis, as they correspond to highly off-resonant processes that have negligible impact on the system dynamics. (We have performed example dynamics calculations for realistic system parameters and initial states, and found that setting these coefficients to zero has no visible effect.) To streamline our demonstration, we choose one specific assignment of effective spin states, corresponding to the scheme where CE spin states have quantum numbers $n',n'+2$. We define the CC spin states as
\[
\ket{\Ua} \equiv \ket{55 \CC_-}, \quad \ket{\Da} \equiv \ket{56 \CC_-},
\]
and define the CE spin states as
\[
\ket{\ua} \equiv \ket{71 \CC_+}, \quad \ket{\da} \equiv \ket{73 \EE_+}.
\]
For this choice of states, the CC and CE transition energies are similar at zero field, making it easy to minimize the F\"orster defect with small fields. We believe that the results below are universal across different choices of quantum numbers $n,n'$.

The below interaction coefficients have been obtained by calculating the Hamiltonian $\hat{H}_{2,\mathrm{eff}}$ for each given $\theta,\Edc$. For each tested value of $\Edc$, the magnetic field $B$ is tuned to $B_\mathrm{res}$, the value that minimizes the F\"orster defect, to make the results more experimentally relevant. We have checked that setting $B$ to different values does not qualitatively alter our findings.

Throughout this analysis, we fix the azimuthal angle $\phi$ to zero and the inter-atomic distance to $| \vec{R}_j - \vec{R}_k | = 7\,\mu\mathrm{m}$. The $\phi=0$ assumption guarantees that all coefficients are real (we do not consider the effects of complex interaction coefficients in this paper). For our configuration, the chosen distance $7\,\mu\mathrm{m}$ makes the dipolar interactions fairly strong, \emph{i.e.} on the order of MHz. The expected radiative lifetime for $N$ atoms in circular states (\emph{i.e.} time until the radiative decay of one of the atoms) is on the order of $\approx \left( 10^{-2}/N \right)\,\mathrm{s}$, so the MHz-level interaction allows for many evolution cycles within the system's operational lifetime. On the other hand, for this interaction strength the spin subspace still remains largely isolated from non-spin states, and we find $\kappa \ge 0.988$ across all cases considered. While not perfect, this level of isolation suggests that leakage effects remain limited, allowing for a substantial number of evolution cycles before significant decoherence occurs.

In the figures, we present interaction coefficients for ${ 0 \le \theta \le \pi/2 }$. However, the full angular dependence for ${ 0 \le \theta \le 2\pi }$ can be inferred by noting that the magnitudes of the coefficients are symmetric about multiples of $\pi/2$.

\subsubsection{Transverse (spin-exchange) interactions}

\begin{figure}
    \centering
    \includegraphics[width=\linewidth]{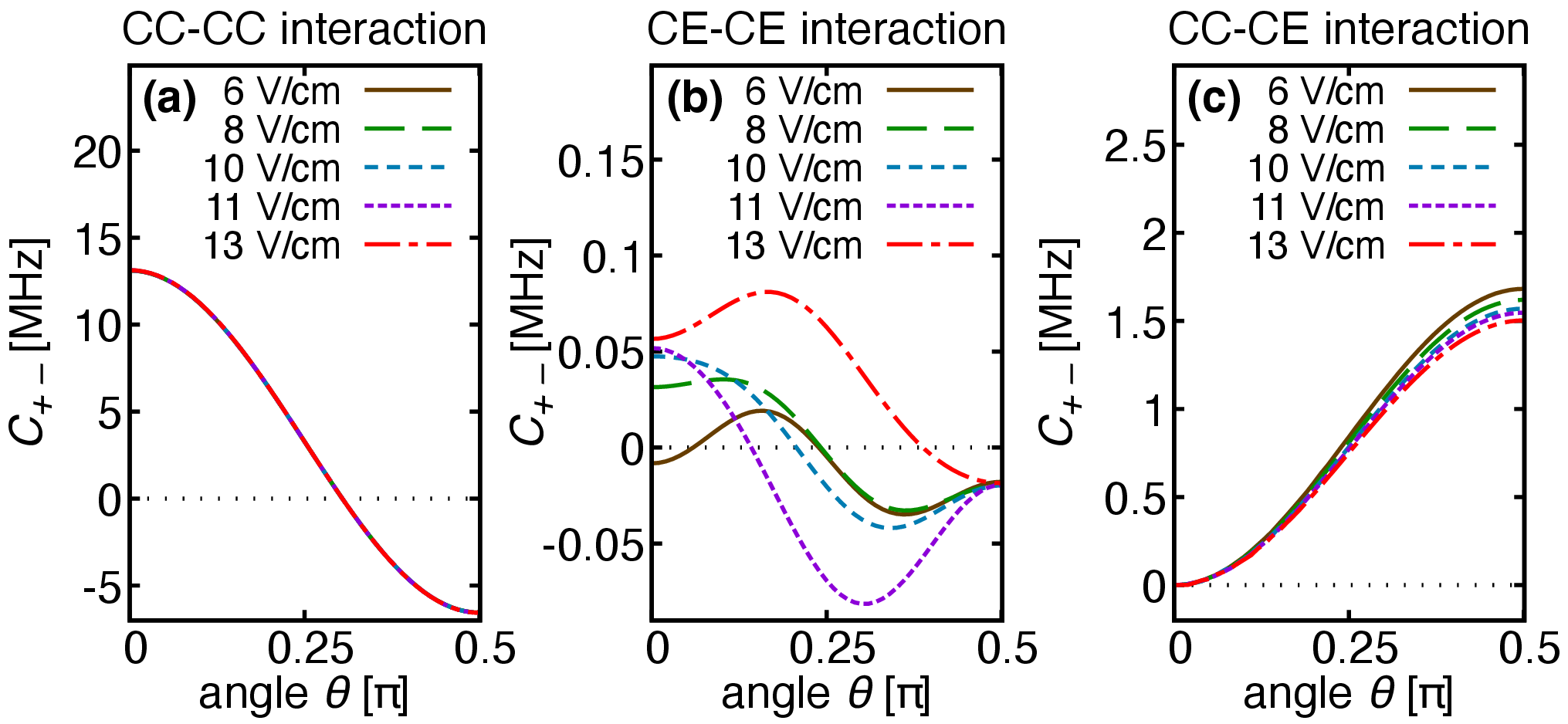}
    \caption{Effective spin-exchange coefficients $C_{+-}$ for two atoms as a function of inter-atomic angle $\theta$ and electric field $\Edc$. The inter-atomic distance is set to $|R| = 7\micron$. (a) $C_{+-}$ for interactions between two CC atoms; curves for different $\Edc$ overlap due to negligible dependence on $\Edc$. (b) $C_{+-}$ for interactions between two CE atoms. (c) $C_{+-}$ for interactions between CC and CE atoms. In all cases, the magnetic field $B$ is set to the value $B_\mathrm{res}$ that minimizes the F\"orster defect. For the five used values of $\Edc$ ($6, 8, 10, 11, 13\,\mathrm{V/cm}$), the values used for $B_\mathrm{res}$ are respectively: $784.07\,\mathrm{G}, 727.82\,\mathrm{G}, 678.36\,\mathrm{G}, 656.33\,\mathrm{G}, 617.97\,\mathrm{G}$.}
    \label{fig:Coefficients_ElectricField_SpinExchange}
\end{figure}

We start by examining the spin-exchange coefficient $C_{+-}$, which is shown in Fig.~\ref{fig:Coefficients_ElectricField_SpinExchange} as a function of $\theta$ and $\Edc$, for different combinations of spin species. 

For a pair of CC spins [Fig.~\ref{fig:Coefficients_ElectricField_SpinExchange}(a)], the coefficient $C_{+-}$ is simply given by the first-order dipolar interaction, \emph{i.e.}, the matrix element ${ \la\Ua\Da|\hat{V}^{(j,k)}\ket{\Da\Ua} }$. Since the total $m_\ell$ of the atom~pair remains unchanged in the exchange, $C_{+-}(\theta)$ depends on the angle as $(1 - 3\cos^2\theta)$ and hence vanishes at the ``magic angle'' $\theta_m \approx 0.304\pi$. Furthermore, $C_{+-}$ is unaffected by the electric field, since the composition of circular states $\ket{\Ua}, \ket{\Da}$ is largely insensitive to $\Edc$. Second- and higher order dipolar contributions are negligible, because there are no pair states that could act as a resonant intermediate state between $\ket{\Ua \Da}$ and $\ket{\Da \Ua}$.

For two CC spins, the magnitude of $C_{+-}$ is proportional to the square of the transition matrix element ${\ket{n \CC_\pm} \leftrightarrow \ket{(n+1) \CC_\pm}}$, and hence it scales as $n^4$. In our configuration, $C_{+-}$ reaches values on the order of $10\,\mathrm{MHz}$.

For the spin exchange between two CE spins [Fig.~\ref{fig:Coefficients_ElectricField_SpinExchange}(b)], the situation is more intricate. Using $\ket{n'\CC}$ and $\ket{(n'+2)\EE}$ as the spin states $\ket{\ua},\ket{\da}$ allows a wider range of higher-order dipolar couplings. In particular, the two pair states $\ket{\ua \da}$ and $\ket{\da \ua}$ can be coupled through intermediate states with quantum numbers $(n'+1),(n'+1)$. As a result, $C_{+-}$ for CE spins includes strong higher-order couplings that modify its angular dependence. Furthermore, in our configuration one of the intermediate states, $\ket{(n'+1)\CC,(n'+1)\CC}$, becomes Stark-resonant with $\ket{\ua \da}$ for $\Edc \approx 11\,\mathrm{V/cm}$. Hence the shape of $C_{+-}$ undergoes a particularly strong change as the electric field is tuned through this value. These higher-order couplings and Stark resonances make $C_{+-}$ highly sensitive to the electric field, and this allows flexible tuning of its angular dependence $C_{+-}(\theta)$. In particular, adjusting $\Edc$ allows to modify the ``magic angle'' $\theta_m$ at which $C_{+-}(\theta_m) = 0$.

The magnitude of $C_{+-}$ for two CE spins is still mainly dictated by the first-order dipolar coupling. This coupling is proportional to the square of the $\ket{n' \CC} \leftrightarrow \ket{(n'+2) \EE}$ transition matrix element, and hence scales only as $(n')^{3}$. Therefore the spin exchange is much weaker than the CC-CC exchange. In our configuration, it is on the order of $10\,\mathrm{kHz}$.

For CC-CE exchange [Fig.~\ref{fig:Coefficients_ElectricField_SpinExchange}(c)], $C_{+-}$ is again given by the first-order dipolar interaction, $\bra{\Da\ua} \hat{V}^{(j,k)} \ket{\Ua\da}$. The second- and higher order contributions are negligible, due to the lack of suitable pair states that are resonant with both $\ket{\Ua\da}$ and $\ket{\Da\ua}$ (even if $\ket{\Ua\da}$, $\ket{\Da\ua}$ are matched in energy via external fields, other pair states $\ket{q}$ are still far in energy from $\ket{\Ua\da}$, $\ket{\Da\ua}$). The $C_{+-}$ magnitude is now a product of strong $\ket{n\CC} \leftrightarrow \ket{(n+1)\CC}$ and weak $\ket{n'\CC} \leftrightarrow \ket{(n'+2)\EE}$ couplings. Hence it scales as $n^2 (n')^{3/2}$, and falls between the values observed for the CC-CC and CE-CE cases. In our configuration, it reaches values on the order of $1\,\mathrm{MHz}$. Additionally, the magnitude is dependent on $\Edc$, due to the strong coupling of the electric field to the elliptical states.

Notably, for CC-CE spin exchange, the angular dependence of $C_{+-}(\theta)$ can be tailored by selecting different values of $m_\ell$ for the effective spin states. For example, in our configuration, the process $\ket{\Ua\da} \leftrightarrow \ket{\Da\ua}$ changes the total $m_\ell$ by 2. Hence the matrix element $\bra{\Da\ua} \hat{V}_{dd} \ket{\Ua\da}$ has the form $\propto \sin^2\theta$ and the spin-exchange interaction vanishes at $\theta_m = 0$.

\subsubsection{Effective energy-shift terms $U$}

\begin{figure}
    \centering
    \includegraphics[width=\linewidth]{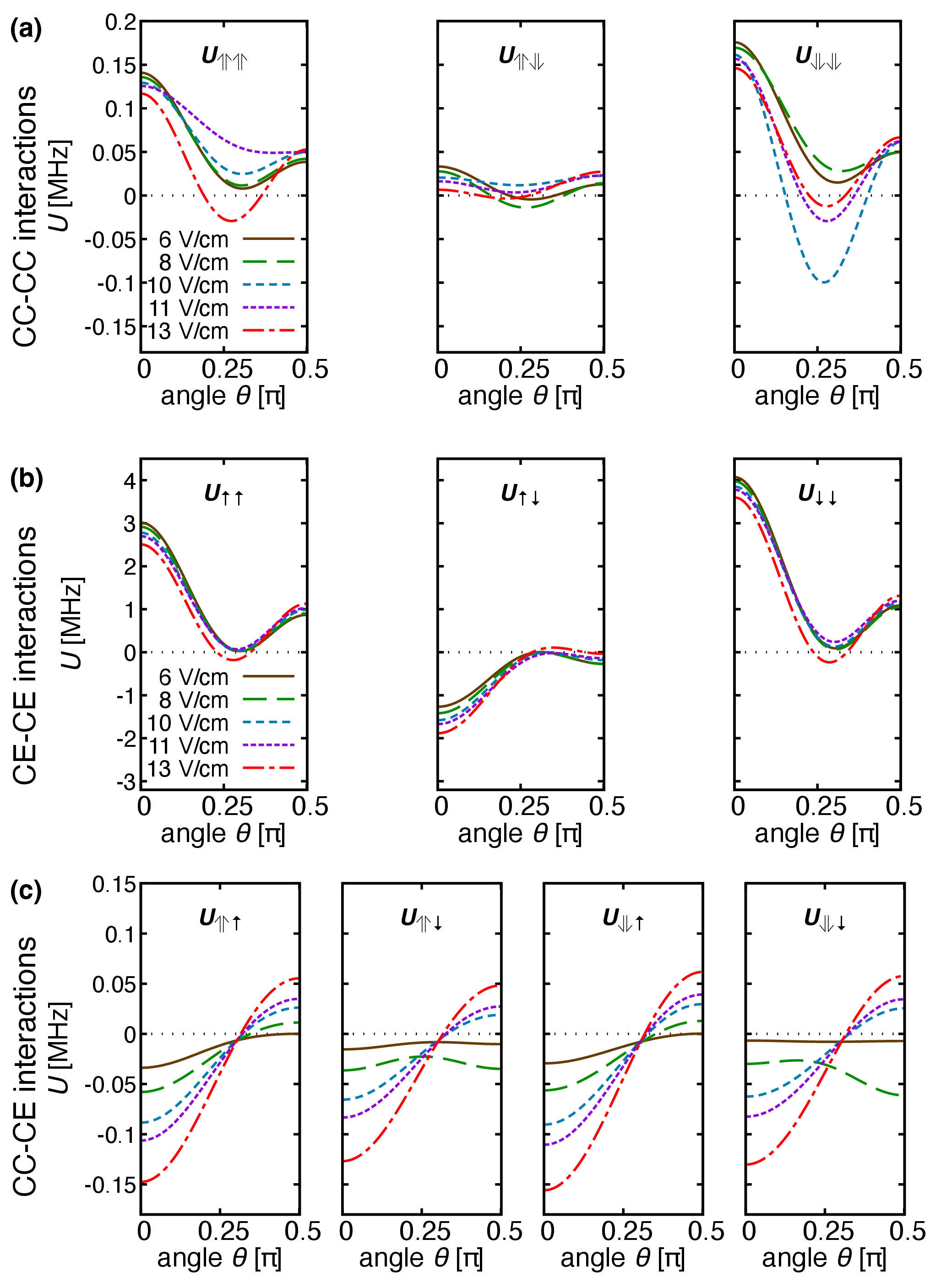}
    \caption{Effective interaction-induced energy shifts $U$ for different effective spin pair states, as a function of $\theta$ and $\Edc$. The magnetic field $B$ is set to $B_\mathrm{res}$ and the inter-atomic distance is $7\,\micron$. (a) $U$ coefficients for two CC atoms: $U_{\Ua \Ua}$, $U_{\Ua \Da}$, $U_{\Da \Da}$ (with $U_{\Da \Ua} = U_{\Ua \Da}$ omitted). (b) $U$ coefficients for two CE atoms: $U_{\ua \ua}$, $U_{\ua \da}$, $U_{\da \da}$ (with $U_{\da \ua} = U_{\ua \da}$ omitted). (c) $U$ coefficients for a CC and a CE atom: $U_{\Ua \ua}$, $U_{\Ua \da}$, $U_{\Da \ua}$, $U_{\Da \da}$.}
    \label{fig:Coefficients_ElectricField_U}
\end{figure}

Next, we consider the diagonal terms $U_{\sigma\sigma'}$, which represent the interaction-induced energy shifts of specific effective spin states $\ket{\sigma,\sigma'}$. These ``$U$ coefficients'' are crucial, as they make up the longitudinal interaction $C_{zz}$ and affect the local effective Zeeman fields $C_z^j$. Figure~\ref{fig:Coefficients_ElectricField_U} shows the $U_{\sigma\sigma'}$ coefficients for different combinations of spin species. 

We first consider the coefficients $U$ for two CC spins [see Fig.~\ref{fig:Coefficients_ElectricField_U}(a)]. Each of these coefficients can be broken down into two main parts. The first is the \mbox{first-order} dipolar interaction between the two atoms, proportional to the field-induced permanent dipole moments of circular states. In our configuration, this contribution is typically on the order of between 10 and 100 kHz. The second part is the second-order contribution, involving the coupling of the spin state $\ket{\sigma,\sigma'}$ to itself via intermediate non-spin states $\ket{q}$. Usually, just one or two states $\ket{q}$ dominate this coupling. For example, the state $\ket{\Ua \Ua} = \ket{n \CC, n \CC}$ is most strongly coupled to $\ket{(n+1) \CC, (n-1) \CC}$ and $\ket{(n-1) \CC, (n+1) \CC}$. These two states dominate the second-order contribution because they are energetically close to $\ket{\Ua \Ua}$ (since the single-atom differences in $n$ compensate each other) and additionally benefit from strong $\CC \rightarrow \CC$ couplings. The total second-order contribution overwhelms the first-order term, and imparts the overall shape ${\propto (1-3 \cos^2 \theta)^2}$ to $U_{\Ua \Ua}(\theta)$. For $U_{\Da \Da}$ the situation is analogous. For $U_{\Ua \Da}$, however, there is only one suitable non-spin state of this kind: $\ket{\Ua \Da} = \ket{n\CC (n+1)\CC}$ couples strongly only to $\ket{(n-1)\CC (n+2)\CC}$. Primarily for this reason, $U_{\Ua \Da} = U_{\Da \Ua}$ is significantly smaller that $U_{\Ua \Ua}, U_{\Da \Da}$.

For two CE spins [see Fig.~\ref{fig:Coefficients_ElectricField_U}(b)] the situation is similar; once again, second-order couplings dominate, and the resulting couplings give the coefficients $U$ a shape close to $\propto (1-3 \cos^2 \theta)^2$. However, there are several important differences. First, and most obviously, the resulting dipolar contributions are stronger by a factor of $\approx (n'/n)^8$ than for the CC case, because of the scaling of the involved transition dipole moments. Second, the coefficient $U_{\ua \da}$ has not one, but two non-spin states $\ket{q}$ contributing strongly to the second-order term: $\ket{\ua \da} = \ket{n'\CC (n'+2)\EE}$ couples strongly to $\ket{(n'+1)\CC (n'+1)\EE}$ and $\ket{(n'-1)\CC (n'+3)\EE}$. This is thanks to the fact that we are using both circular and elliptical states: it means the non-spin subspace has a larger variety of pair states which can couple strongly and resonantly to $\ket{\ua \da}$. However, for these two $\ket{q}$ the corresponding energy differences have opposite signs, and the net coefficient $U_{\ua \da}$ has the opposite sign to $U_{\ua \ua}$ and $U_{\da \da}$. This sign reversal will turn out to have significant effects for the spin-spin interactions.

Finally, we examine the interspecies CC-CE interaction [Fig.~\ref{fig:Coefficients_ElectricField_U}(c)]. This time, the second-order couplings are much weaker. This is because the involved atom states have significantly different principal quantum numbers $n, n'$, which reduces the resonance with the non-spin states. For example, for the state $\ket{\Ua \ua} = \ket{n\CC n'\CC}$, the strongest second-order contributions involve the two states $\ket{(n\pm1) \CC (n'\mp 1)\CC}$. However, because $n,n'$ have very different values, the single-particle energy differences do not compensate. The resulting off-resonance between pair states greatly reduces the second-order contribution. Additionally, the two states contribute with opposite signs but very similar magnitudes, and their contributions nearly cancel out. The net result is that all the coefficients $U$ for CC-CE atom~pairs are dominated by the first-order coupling. The resulting $U$ coefficients have the shape ${ U(\theta) \propto (1-3 \cos^2 \theta) }$ characteristic for diagonal dipolar interaction, especially at higher $\Edc$, where the interaction between permanent dipoles is stronger. Additionally, all four $U$ now have similar magnitudes and signs. This will be crucial for the atomic interaction.

\subsubsection{Local Zeeman field acting on individual spins}

\begin{figure}
    \centering
    \includegraphics[width=\linewidth]{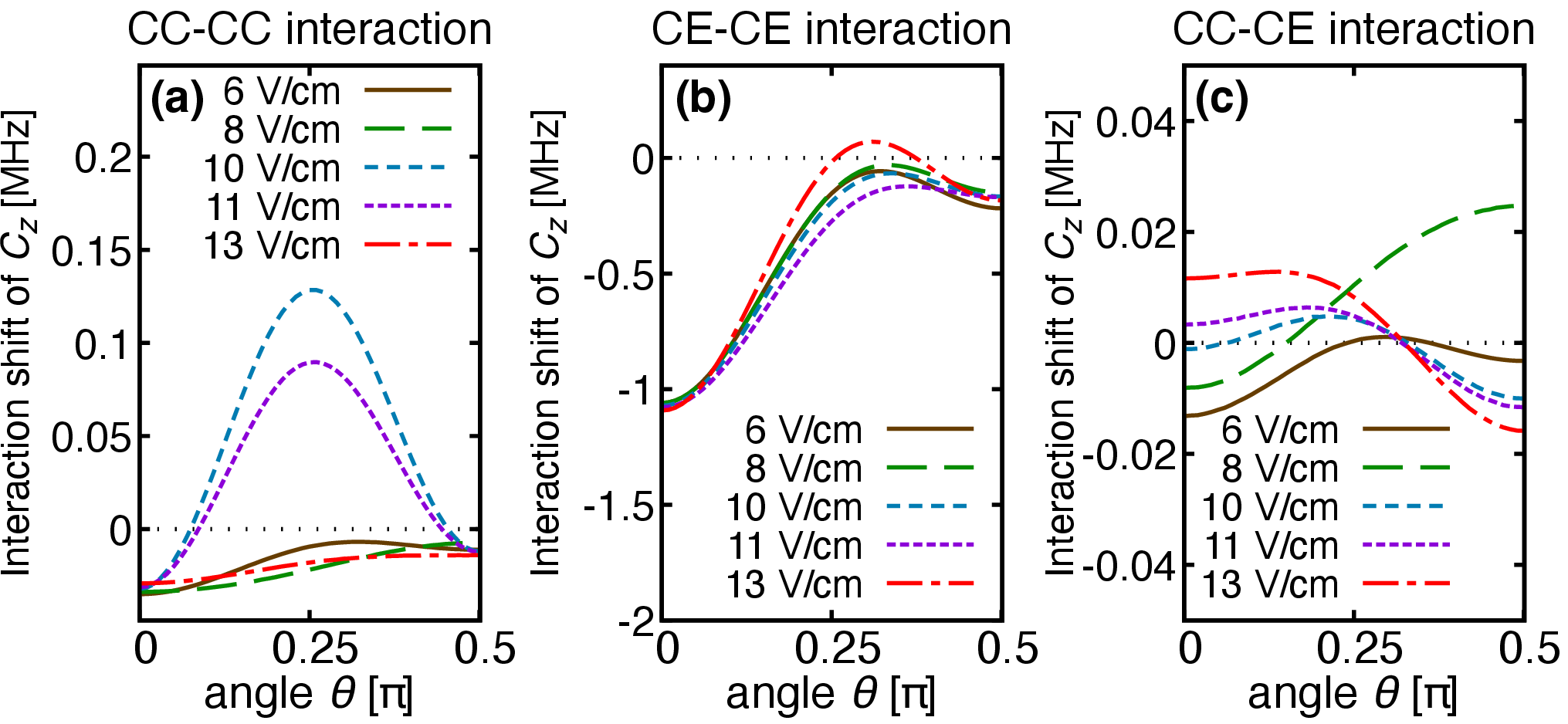}
    \caption{The interaction shift of $C_z^j$ [Eq.~\eqref{eq:cjz_int_contribution}] for the atom $j$, from one other atom placed at a distance of 7 micrometers. (a) Contribution for a CC atom, from a nearby CC atom. (b) Contribution for a CE atom, from a nearby CE atom. (c) Contribution for a CC atom, from a nearby CE atom.}
    \label{fig:Coefficients_ElectricField_CjzShift_diffEdc}
\end{figure}

We now analyze the local effective Zeeman field $C_z^j$ acting on individual spins. $C_z^j$ has been given earlier in Eq.~\eqref{eq:coefficients} as

\begin{subequations}
    \label{eq:cjz_definition}
    \begin{align}
        C_z^j &= \left( \bra{\ua^j} \hat{h}^{(j)} \ket{\ua^j} - \bra{\da^j} \hat{h}^{(j)} \ket{\da^j} \right) \label{eq:cjz_1body_contribution}\\
        &\quad+ \sum_{k \ne j} \frac{U^{jk}_{\uparrow\uparrow} + U^{jk}_{\uparrow\downarrow} - U^{jk}_{\downarrow\uparrow} - U^{jk}_{\downarrow\downarrow}}{2}. \label{eq:cjz_int_contribution}
    \end{align}
\end{subequations}

The first term (Eq.~\eqref{eq:cjz_1body_contribution}) corresponds to the energy difference between the two $n$ manifolds corresponding to pseudo-spin states ($n, n+1$ or $n', n'+2$). For Rydberg states this difference is on the order of $10\,\mathrm{GHz}$. It is usually the largest energy scale in the system. In particular, it dictates the energy cost of changing the system's total spin. Our calculations (not shown here) indicate that the non-spin-conserving coefficients \( C_+ \), \( C_{z+} \), and \( C_{++} \) are much smaller than this energy scale. Specifically, \( C_{++} \) for each spin species combination is comparable in magnitude to \( C_{+-} \) (up to approximately 10 MHz), while \( C_+ \) and \( C_{z+} \) reach at most 1 MHz. Therefore the non-spin-conserving processes are highly off-resonant and have a negligible effect on the system's dynamics.

The second term (Eq.~\eqref{eq:cjz_int_contribution}) is an interaction-induced shift from neighboring atoms. This term can influence spin-exchange dynamics if it varies across different atoms by amounts on the order of $C_{+-}$. To examine this possibility, in Fig.~\ref{fig:Coefficients_ElectricField_CjzShift_diffEdc} we plot the value of Eq.~\eqref{eq:cjz_int_contribution} for a single pair of atoms $j,k$ at angle $\theta$. This gives the idea of how much $C_z^j$ may vary depending on the surroundings of atom $j$.

The resulting contributions to $C^j_z$ reflect the magnitude of their comprising coefficients $U$. For atom $j$ of species CC, a single CC neighbor [Fig.~\ref{fig:Coefficients_ElectricField_CjzShift_diffEdc}(a)] or a CE neighbor [Fig.~\ref{fig:Coefficients_ElectricField_CjzShift_diffEdc}(c)] induces a shift in $C_z^j$ ranging from $0$ to $150\,\mathrm{kHz}$. Considering that $C_{+-}$ for CC-CC interactions is $\sim 10\,\mathrm{MHz}$, this shift is comparatively small. Therefore, in a spatially uniform atom distribution, where each atom sees roughly the same number of neighbors, fluctuations in $C_z$ across different atoms will be small and will have minimal effect on spin exchange.

For atom $j$ of species CE, a single CC neighbor induces a $C^j_z$ shift up to $150\,\mathrm{kHz}$ [not pictured in Fig.~\ref{fig:Coefficients_ElectricField_CjzShift_diffEdc}], similarly as the opposite case. A single CE neighbor [Fig.~\ref{fig:Coefficients_ElectricField_CjzShift_diffEdc}(b)] induces a shift in $C_z^j$ up to $1\,\mathrm{MHz}$. These shifts are significantly larger than the spin-exchange energy scales $C_{+-}$ for CE-CE interactions ($\approx 10\,\mathrm{kHz}$). This implies that in a system with CE atoms, variations in $C_z^j$ could inhibit spin-exchange interactions between these atoms, leading to localization effects.

\subsubsection{The longitudinal interaction coefficients}

\begin{figure}
    \centering
    \includegraphics[width=\linewidth]{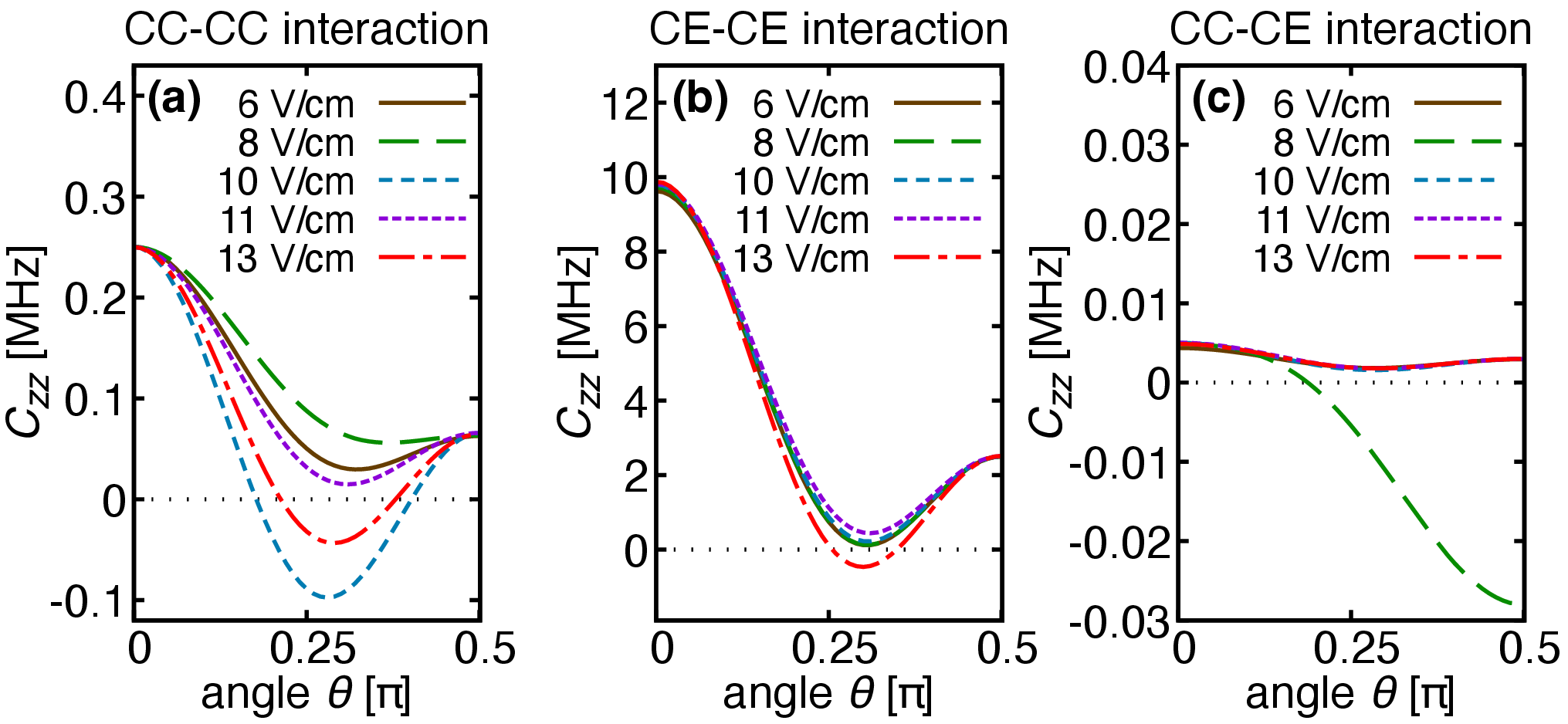}
    \caption{Effective longitudinal spin interaction coefficients $C_{zz}$ between two atoms as a function of $\theta$ and $\Edc$. The magnetic field $B$ is set to $B_\mathrm{res}$ and the inter-atomic distance is $7\,\micron$. (a) $C_{zz}$ for two CC atoms. (b) $C_{zz}$ for two CE atoms. (c) $C_{zz}$ for interactions between CC and CE atoms.}
    \label{fig:Coefficients_ElectricField_Longitudinal}
\end{figure}

The longitudinal interaction coefficients $C_{zz}$, defined by
\begin{equation}
\label{eq:sum_of_u_giving_Czz}
C^{jk}_{zz} = U^{jk}_{\ua\ua} - U^{jk}_{\ua\da} - U^{jk}_{\da\ua} + U^{jk}_{\da\da},
\end{equation}
are depicted in Fig.~\ref{fig:Coefficients_ElectricField_Longitudinal} for various spin species combinations. 

For CC-CC interactions [Fig.~\ref{fig:Coefficients_ElectricField_Longitudinal}(a)] and CE-CE interactions [Fig.~\ref{fig:Coefficients_ElectricField_Longitudinal}(b)], the angular dependence of $C_{zz}$ reflects that of the underlying $U$ coefficients, and it is approximately $\propto (1-3 \cos^2 \theta)^2$ for both CC-CC and CE-CE interactions. However, the magnitudes differ starkly. For CC spins, $C_{zz}$ is generally small compared with $C_{+-}$. Therefore, the CC-CC spin interactions are approximately of Heisenberg type ($C_{+-} \gg C_{zz}$). For CE spins, $C_{zz}$ greatly exceeds $C_{+-}$; this is both due to the larger $U$ values from the higher $n'$, and due to the opposite signs of $U^{jk}_{\ua\da},U^{jk}_{\da\ua}$ vs. $U^{jk}_{\ua\ua},U^{jk}_{\da\da}$, which cause all terms in Eq.~\eqref{eq:sum_of_u_giving_Czz} to add constructively. Therefore, two CE spins exhibit Ising-like interactions ($C_{zz} \gg C_{+-}$).

For interactions between CC and CE spins [Fig.~\ref{fig:Coefficients_ElectricField_Longitudinal}(c)], $C_{zz}$ is very small. This is because the contributions from $U$ coefficients nearly cancel out, given their similar magnitudes and signs. In our configuration, the resulting $C_{zz}$ is on the order of $10\,\mathrm{kHz}$, which can be essentially neglected when compared with other energy scales in the system. Therefore, the CC-CE spin interactions are ultimately of Heisenberg type ($C_{+-} \gg C_{zz}$).

\subsubsection{Summary of the results}

Let us summarize the characteristic properties of the interactions between CC and CE spins.

The three combinations of interacting species correspond to different interaction types. CC-CC interactions feature strong transverse interaction ($S_+ S_-$) and weak longitudinal interaction ($S_z S_z$), oppositely to CE-CE interactions, which exhibit weak transverse interaction and strong longitudinal interaction. The CC-CE interspecies interactions have a moderately strong transverse component, and a near-vanishing longitudinal component. Therefore, in general, CC-CC and CC-CE interactions are of Heisenberg-like type ($C_{+-} \gg C_{zz}$) while CE-CE interactions are of Ising-like type ($C_{+-} \ll C_{zz}$). Additionally, unlike for CC spins, the spin exchange between CE spins can be strongly affected by local shifts of $C_z^j$ caused by interactions with the local environment.

It is worth noting that, beyond spin model simulations, this setup offers intriguing possibilities for simulating Hubbard lattice models. A system of fixed 1/2-spins can be mapped onto that of hardcore bosons hopping between lattice sites~\cite{1956-Matsubara-ProgTheorPhys}. In this picture, the atoms represent lattice sites, and the two ``pseudospin levels'' instead represent different states of a lattice site---occupied or vacant. The effective spin-exchange between two atoms is reinterpreted as the hopping of a particle between sites. The ``two-species'' approach allows to represent two distinct sublattices, such that the couplings are of different character within and between each sublattice. This allows to explore variants of topological lattice models, such as the Su-Schrieffer-Heeger (SSH) model~\cite{1979-Su-PRL}, the Rice-Mele model~\cite{2019-Cooper-RevModPhys}, the Hofstadter-Harper model~\cite{2013-Aidelsburger-PRL}, and the Kitaev model~\cite{2008-Baskaran-PRB}.

Integrating both spin species in a single simulated model provides significant benefits over a single-species setup, both for spin models and Hubbard lattice models. First, a two-species approach offers greater flexibility in designing interactions between different parts of the system. For example, it allows to easily divide a simulated lattice model into two sub-lattices with independent properties. Second, the broader range of ``magic angles'' accessible with the two-species setup gives more flexibility in designing the geometry of the simulator. For example, a spin chain with suppressed transverse interactions can be achieved using a line of CC spins set at the specific angle $\theta = 0.304\pi$, but it can also be realized with CE spins set at almost any polar angle, enabling new designs for the simulation array. Third, the CE spins allow more precise tuning of Ising-like interactions, which for CC spins are restricted to a narrow range near $\theta = 0.304\pi$ and are highly sensitive to atom positioning. These combined advantages make the dual-species approach more robust and versatile.

\section{Example simulated systems}\label{sec:example-systems}

To illustrate the potential applications of our setup, we describe two examples of realizable system geometries. These examples will also demonstrate the advantages of combining the CC and CE spins, compared with CC spins alone.

\begin{figure}
    \centering
    \includegraphics[width=0.9\linewidth]{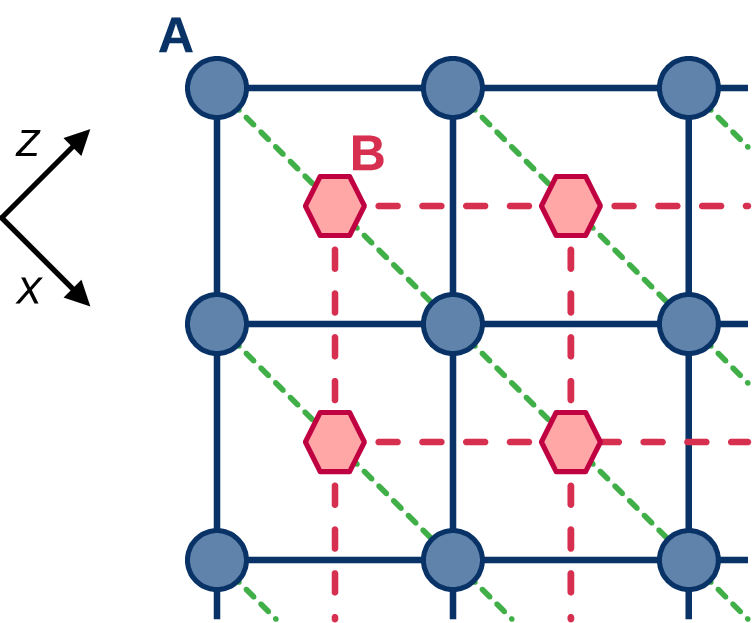}
    \caption{Example geometry realizable with the two-species approach: A two-dimensional atom arrangement consisting of two interleaved square lattices, labeled ``A'' and ``B,'' each containing atoms of opposite spin species. The blue circles and red hexagons stand for atoms of the two different spin species, CC and CE (the specific correspondence between shapes and species is arbitrary). When the quantization axis (axis $Z$) is oriented at an angle $\pi/4$ relative to the unit cell sides, the interactions between neighboring atoms within each lattice, shown as blue solid lines (lattice ``A'') and red long-dashed lines (lattice ``B''), become uniform. However, the inter-lattice interactions (green short-dashed lines) display spatial anisotropy, only acting along one direction.}
    \label{fig:two-species-example_systems-figure_1}
\end{figure}

\begin{figure}
    \centering
    \includegraphics[width=1.0\linewidth]{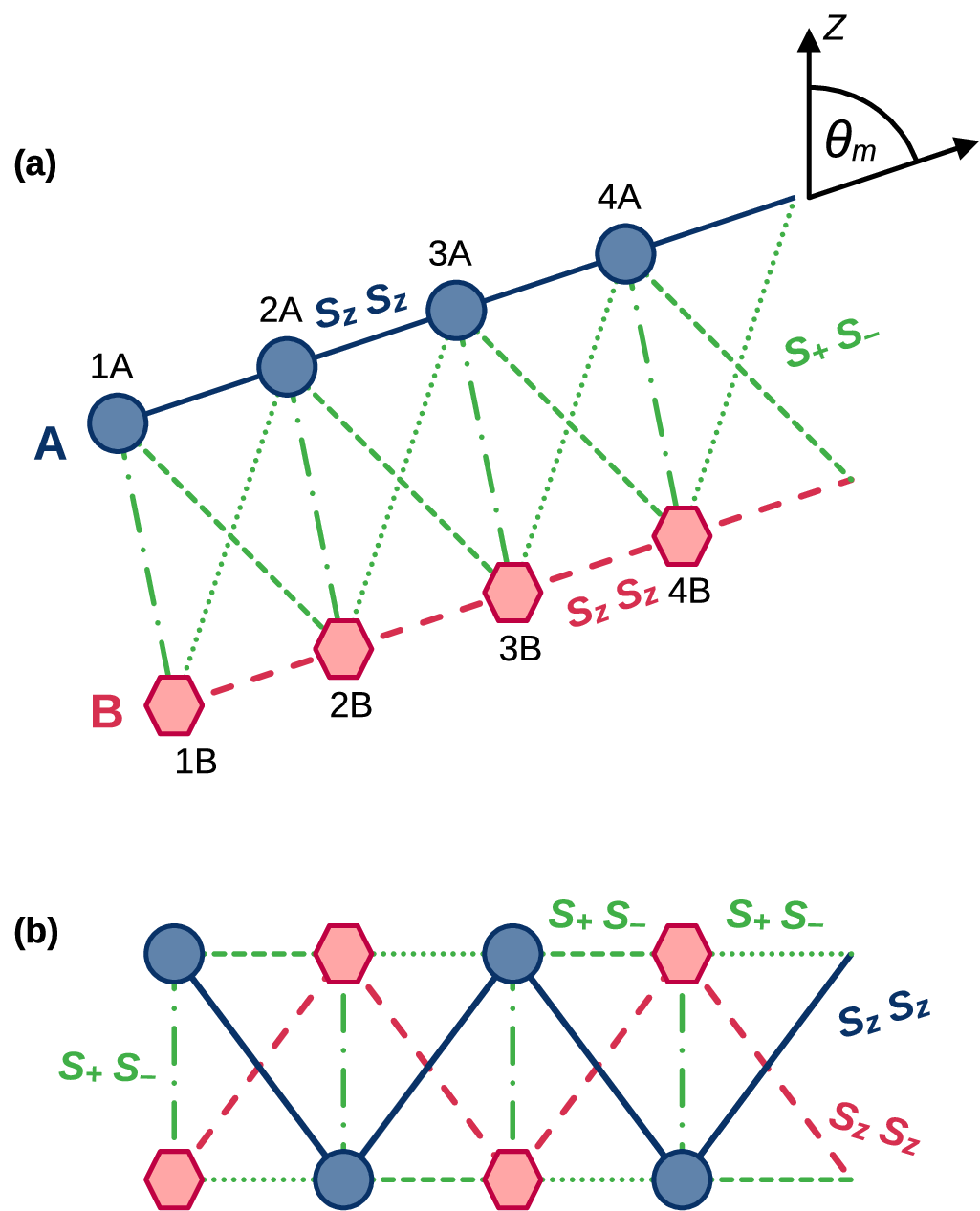}
    \caption{Example geometries realizable with the two-species approach. The blue circles and red hexagons stand for atoms of the two different spin species, CC and CE (the specific correspondence between shapes and species is arbitrary). (a) A two-dimensional arrangement forming two parallel spin chains composed of different species, aligned along the magic angle $\theta_m = 0.3041\pi$. This configuration allows to suppress spin-exchange interactions for both CC and CE atoms, resulting in two chains with purely Ising $S_z S_z$ couplings (blue solid and red dashed lines) with differing strengths. The particles in each chain are labeled as $1A, 2A, 3A, \ldots$ and $1B, 2B, 3B, \ldots$. The inter-chain interactions (green lines) are spin-exchanging. Only the three strongest interactions--those between each atom and its three nearest neighbors of the opposite chain--are shown. These couplings vary in strength depending on their relative angles, indicated by distinct line styles. (b) An alternate representation of the system in (a) without considering physical positioning. The system can be viewed as two chains with alternating sites $1A, 2B, 3A, 4B, \ldots$ and $1B, 2A, 3B, 4A, \ldots$. Each chain features spin-exchanging interactions of alternating strengths (green dashed and dotted lines), while the two chains are coupled by Ising interactions (blue solid and red dashed lines) and interspecies spin-exchange interactions (green dash-dotted lines).}
    \label{fig:two-species-example_systems-figure_2}
\end{figure}

As the first example, we present a ``double lattice'' model, illustrated in Fig.~\ref{fig:two-species-example_systems-figure_1}. This model consists of spins arranged in a periodic 2D square lattice (lattice ``A''), with an additional spin at the center of each unit cell, forming a secondary lattice ``B''. Such ``double lattice'' configurations find application in multiple branches of physics, such as research on photonic crystals (nanostructures with spatially periodic dielectric functions)~\cite{2002-Malkova-PRB} or multi-layer materials~\cite{2013-Bjorkman-SciRep,2020-Leconte-2DMat}. In the presented system, spins within each sublattice (``A'' and ``B'') interact with their four nearest neighbors with equal magnitudes of couplings, although the interaction strength differs between the two sublattices. Furthermore, the model includes ``A''-``B'' couplings that are spatially anisotropic, acting exclusively along one direction within a unit cell.

The model in Fig.~\ref{fig:two-species-example_systems-figure_1} can be readily engineered using a combination of CC and CE species, by assigning different spin species to the two lattices and arranging the atoms into a square lattice structure at a $\pi/4$ angle relative to the quantization axis. Consequently, the nearest-neighbor interactions within the same lattice have identical magnitudes in all four directions, while the nearest-neighbor interactions between different lattices vanish for atom~pairs set at the magic angle $\theta = 0$. This system would be difficult to engineer with CC spins alone, as it would be challenging to simultaneously achieve inter-lattice interactions that only act in one direction, and within-lattice interactions that act with equal strength in all directions.

Another example is shown in Fig.~\ref{fig:two-species-example_systems-figure_2}(a). This model represents a spatially periodic 2D structure, composed of a pair of Ising chains, labeled ``A'' and ``B''. Each chain features uniform $S_z S_z$ nearest-neighbor couplings, but with different interaction strength for chains ``A'' and ``B''. Additionally, each particle $jA$ in chain ``A'' undergoes spin-exchanging interactions with its opposite-chain partner $jB$ and its two nearest neighbors, $(j \pm 1)B$ (interactions with further neighbors are smaller by about an order of magnitude, and thus are typically unimportant). This system can be realized using CC and CE spins in the following manner: by assigning different spin species to each chain, aligning both chains along the magic angle $\theta_m = 0.3041\pi$ to nullify the CC-CC spin-exchange interactions, and adjusting the external electric field so that the CE-CE spin-exchange interactions also vanish at $\theta_m$. The spin-exchanging interspecies interactions arise naturally once the chains are placed close to each other.

Note that engineering this system using only CC spins would pose problems. While two Ising chains could be created by aligning them along angle $\theta_m$, achieving different interaction strengths for both chains would be difficult. One would either have to use different inter-particle distances in the two chains (which would destroy the periodicity of the inter-chain interactions), or use circular states with different $n$ for the two chains (in which case, it would not be possible to tune the transitions into resonance, precluding spin exchange between the chains).

By ignoring the physical positioning and treating particles $jA, (j+1)B, (j+2)A, \ldots$ as a single chain, the system in Fig.~\ref{fig:two-species-example_systems-figure_2}(a) can be made equivalent to another model: two identical chains where each chain has spin-exchange interactions of alternating strengths, as shown schematically in Fig.~\ref{fig:two-species-example_systems-figure_2}(b). These chains are connected to each other by both $S_z S_z$ and spin-exchanging couplings. 

The model in Fig.~\ref{fig:two-species-example_systems-figure_2}(b) can be also viewed in a different light by mapping the atoms to lattice sites and spin states to site occupancy. Then, each of the two chains can be re-interpreted as a Su-Schrieffer-Heeger (SSH) lattice, a fundamental topological model characterized by alternating hopping amplitudes between even and odd sites. The system can also be seen as a variant of the two-leg Creutz ladder, another well-known topological system~\cite{2017-Junemann-PRX}. In the lattice picture, the transverse spin interactions on each chain are equivalent to the hopping of hardcore bosons between sites. Meanwhile, the inter-chain longitudinal interactions ($\hat{S}^i_z \hat{S}^j_z$) can be understood as longitudinal interactions between bosons occupying sites $i,j$. This can be seen by applying the mapping $\hat{S}^i_z \rightarrow (1/2 - \hat{n}_i)$, where $\hat{n}_i$ is the operator measuring the population on site $i$ ($\langle \hat{n}_i \rangle = 0,1$); in this case, the $\hat{S}^i_z \hat{S}^j_z$ term leads to effective $\propto \hat{n}_i \hat{n}_j$ boson-boson interactions. 

Note that in the Fig.~\ref{fig:two-species-example_systems-figure_2}(a) example, the relative positions of the chains can be so engineered that particles $jA, jB$ lie on a line with a polar angle $\theta = 0$. In that case, spin-exchanging interactions between $jA, jB$ [represented by the dash-dotted green lines in Fig.~\ref{fig:two-species-example_systems-figure_2}(a,b)] disappear. Then, the two SSH chains in Fig.~\ref{fig:two-species-example_systems-figure_2}(b) behave approximately as two separate sub-systems with separately conserved populations.


\section{Experimental feasibility}\label{sec:experimental}

\subsection{Trapping}

The implementation of the setup relies on the ability to trap individual Rydberg atoms at specific positions. In experiments using ultracold alkali-metal atoms in their ground states, red-detuned optical tweezers are typically used to trap the atoms at desired locations. However, these tweezers become repulsive for atoms excited to Rydberg states, due to the strong ponderomotive potential of the Rydberg electron~\cite{2000-Dutta-PRL}. Consequently, experiments involving Rydberg atoms require alternative trapping methods. 

Currently, the standard method of optically trapping Rydberg atoms uses ``bottle-beam'' hollow traps. Atoms are confined in an intensity minimum at the center of a laser beam, which is created using a spatial light modulator~\cite{2020-Barredo-PRL}. In a recent experiment, this technique was successfully used to create an array of individually trapped rubidium atoms in circular states~\cite{2023-Ravon-PRL}.

For alkaline-earth atoms, other trapping methods are possible. Unlike for alkali-metal atoms, the ionic core of the alkaline-earth atom remains optically active even in Rydberg states, due to the second valence electron remaining close to the core. This allows to trap circular-state atoms in conventional optical tweezers, as the ionic core's interaction with the tweezer overpowers the contribution from the ponderomotive potential of the Rydberg electron. Tweezer trapping has been experimentally demonstrated for low-$\ell$ Yb Rydberg states~\cite{2022-Wilson-PRL} and, very recently, for circular states of Sr~\cite{2024-Holzl-PRX}.

For elliptical states, to our knowledge, no explicit experimental realizations of trapping few-atom systems have been reported. However, the techniques developed for circular-state trapping are likely adaptable to elliptical-state atoms with minimal modifications.

\subsection{Influence of motional degrees of freedom}

For simplicity, we treat the atoms as fixed points throughout this paper. However, in reality atomic motion within the traps can significantly impact the system's dynamics. For traps with low frequency $\omega$, dipole-dipole interactions between atoms can be strong enough to excite higher motional states. The resulting fluctuations in $\vec{R_i}-\vec{R_j}$ can lead to entanglement between the atomic motion and the internal-level dynamics, equivalent to spin-motion coupling, which induces unwanted dephasing effects. This interaction-induced motion coupling has been demonstrated via numerical calculations for circular-state atoms in Ref.~\cite{2023-Mehaignerie-PRA}.

Thermal excitations of atomic motion can cause similar issues. Ref.~\cite{2024-Mehaignerie-Arxiv} observed significant positional oscillations (up to $1\,\mathrm{\mu m}$) in a setup with two circular-state atoms under typical conditions (trap frequency $\omega=22.8\,\mathrm{kHz}$; interaction strength $\approx 1\,\mathrm{MHz}$ at a distance of $13\,\mathrm{\mu m}$; room temperature). To mitigate these dephasing effects during long-time evolution, our setup should operate at cryogenic temperatures (a few $\mathrm{\mu K}$ or lower), and trap frequencies should be as high as possible.

\subsection{Preparing circular and elliptical states}

Carrying out simulations of specific scenarios requires not only preparing atoms in precisely fixed positions, but also preparing each individual atom in a desired circular or elliptical state, rapidly and reliably.

A number of methods exist that allow rapid excitation of circular levels. One well-established approach is the adiabatic rapid-passage method~\cite{1983-Hulet-PRL,2020-Cortinas-PRL}. The atoms, driven into low-$|m_\ell|$ states, are subjected to a radio-frequency ac field, combined with parallel magnetic and electric fields. Under the combined effects of these fields, the atomic spectrum exhibits an avoided crossing, connecting the low-$|m_\ell|$ state to a circular state of the same $n$. This crossing can then be traversed by adiabatically tuning the electric field. 

A faster variant of the adiabatic rapid-passage method has been proposed as well~\cite{2017-Signoles-PRL}. This variant exploits the fact that, for higher $\ell$, transitions between states with successive $\ell$ are all of nearly the same frequency. In the procedure, the electric field is kept constant, at a value where this transition frequency is on resonance with the chosen radio-frequency radiation. A radio-frequency pulse is then applied for a precise time, driving Rabi oscillations between the initial state (low $|m_\ell|$) and the circular state. After a $\pi$ pulse, the population is coherently transferred into the circular state.

Another well-established approach is the crossed-fields method. Here the atoms are initially placed in perpendicularly crossing electric and magnetic fields, and excited into $m_\ell = 0$ states of the desired $n$. The electric field is then adiabatically switched off, while the transverse magnetic field remains in place. This process transfers the atoms into circular states $\ket{n \CC}$. Note, however, that in our setup this method poses complications: once the circular state preparation is finished, it is still necessary to turn on an electric field parallel to the magnetic field, which needs to be carefully done to avoid disturbing the circular excitations~\cite{1989-Cheret-EPL,1997-Lutwak-PRA,2018-Morgan-PRA}.

Other fast circularization methods involve using quantum control theory, \emph{i.e.} designing specially-shaped radio-frequency pulses that circularize the atomic array with optimal speed and fidelity~\cite{2018-Patsch-PRA,2020-Larrouy-PRX}.

The excitation of elliptical states, which are a key aspect of our setup, is a less developed topic compared with circular state excitation. However, several plausible approaches exist. A straightforward method is to produce atoms in $\ket{(n-1) \CC}$ states, then transfer them to $\ket{n \EE}$ states via a $\pi$-polarized microwave transition. It is also likely that existing circularization schemes can be adapted to produce elliptical states. For example, $\ket{n \EE}$ states can be produced with a high fidelity via optimally shaped pulses~\cite{2024-Huls-Personal}.

Another possibility is a proposed ``quantum Zeno dynamics'' scheme~\cite{2014-Signoles-NatPhys}. The atom is prepared in a $\ket{n \CC}$ state, then a radio-frequency field is applied with a frequency equal to the transition between the different high-$\ell$ levels $\ket{n, \ell}$. This makes the atom state evolve back and forth through the entire ladder of Rydberg levels. By applying a microwave transition to couple a level $\ket{n,\ell'}$ to $\ket{n-1,\ell'}$, splitting that level into two, the atom state is ``blocked'' from evolving past $\ket{n,\ell'}$. In this way, one could prepare the elliptical state for given $n$, or even a linear combination of circular and elliptical states.

\subsection{Atom-resolved state preparation}

Another key requirement of our proposal is the ability to address individual atoms in the setup, in order to initialize each atom into the desired state. Most methods of circular state preparation rely on radio-frequency radiation, which has a large wavelength (on the order of milimeters or above) and thus cannot be locally addressed in systems of Rydberg atoms, where interparticle distances are measured in microns. 

Several proposed approaches address this difficulty by using optical-frequency beams that are addressable on the micron scale. The proposals in Refs.~\cite{2021-Cohen-PRXQ, 2020-Cardman-PRA} use superpositions of copropagating optical Laguerre-Gauss beams, which can drive transitions with a tunable frequency (even non-optical) and very large angular momentum transfer. This enables direct excitation of individual atoms from low-$\ell$ Rydberg states to circular states. Another approach~\cite{2014-Labuhn-PRA} uses optical-frequency addressing beams during the initial Rydberg excitation, to temporarily make specific atoms off-resonant with the excitation beam.

Another possibility opens for alkaline-earth Rydberg atoms. In these atoms, the circular state of the valence electron experiences an energy shift when the ionic atom core is optically excited. This allows optical-wavelength, site-resolved control of the circular states~\cite{2022-Muni-NatPhys,2024-Wirth-PRL}.

Other approaches leverage ancilla atoms to utilize the Rydberg blockade effect, where interactions between two Rydberg atoms shift the energy levels, making simultaneous excitation off-resonant and preventing the excitation of both atoms at once. For example, Ref.~\cite{2023-Mehaignerie-Thesis} describes a simple scheme where a two-atom $\ket{\Uparrow \Downarrow }$ state is prepared by using a third ancilla atom. Depending on its position, the ancilla induces an angle-dependent Rydberg blockade interaction with a nearby atom, which thus becomes off-resonant with the excitation field. By moving the ancilla, the two atoms can therefore be excited into the desired states one at a time.

For certain system geometries, it is possible to use more specialized methods of initial state preparation. For example, in a spin chain of regularly-spaced atoms, one can use optical tweezers to temporarily displace the neighbors of the atom of interest. This shifts the resonant frequencies of the atom, and its state can then be controlled individually~\cite{2016-Nguyen-Thesis}.

\subsection{State detection}

Tracking the simulator state requires state detection on the level of individual atoms, with the ability to distinguish different circular and elliptical states. Optical detection allows to resolve the presence of atoms on individual sites, but in general it does not allow discerning different internal levels. It can, however, be combined with other methods, which rely on removing atoms of specific levels from the array, followed by optical detection to find out which atoms are now missing. This allows a simple way of site-resolved state detection ~\cite{2023-Ravon-PRL}. 

A frequent approach to many-body state readout is the field ionization method: the electric field is gradually increased to cross the ionization thresholds of successive atomic levels. By counting the emitted ions as a function of the ionizing field, and matching specific atomic states to their predicted threshold fields, populations of individual states can be measured. This method is widely applied experimentally to detect circular Rydberg states~\cite{1983-Hulet-PRL,1986-Liang-PRA,2020-Teixeira-PRL,2020-CantatMoltrecht-PRR}.

By itself, the ionization method allows to measure the overall population, but not to detect the states on selected sites. This limitation can be overcome, \emph{e.g.}, by selectively ionizing atoms in one of the states, then imaging the atomic array to identify which atoms were ionized ~\cite{2015-Barredo-PRL,2023-Ravon-PRL}. Another limitation of the ionization method is that it generally cannot distinguish $\ket{n \CC}$ and $\ket{n \EE}$ states, which have very similar threshold ionization fields ~\cite{2020-CantatMoltrecht-PRR}. However, this issue can also be overcome, \emph{e.g.}, by first transferring the elliptical-state atoms to a level from a different energy manifold ~\cite{2014-Signoles-NatPhys,2018-Dietsche-Thesis}.

Other atom-resolved state detection methods are also possible. For instance, non-destructive state measurements can use ancilla atoms that become resonant or non-resonant with a driving field, depending on the state of a nearby array atom~\cite{2021-Cohen-PRXQ}.

Specialized methods of atom-resolved state detection can also be applied in specific configurations. For example, Ref.~\cite{2016-Nguyen-Thesis} considers a one-dimensional spin chain trapped in a lattice, bordered on both ends by ``plug'' beams. To detect the state of the chain atom-by-atom, the lattice is switched off and the plug beam is slowly reduced, expelling atoms one by one from the right-hand end of the chain. This allows to individually read out the atomic states, \emph{e.g.}, by field ionization.

For alkaline-earth Rydberg atoms, another detection method is available, which makes use of the electrostatic coupling between the two valence electrons. The ionic core electron can be driven into a fluorescing state, whose excitation probability depends on the state of the Rydberg electron. By measuring the fluorescence across the system, it is then possible to measure the Rydberg electron state in a quantum-nondestructive way~\cite{2022-Muni-NatPhys}.

\subsection{Maximum lifetimes}

Radiative lifetimes of Rydberg atoms are typically the dominant limitation of the simulation time. Long radiative lifetimes are the largest advantage of using circular states. Circular states have only one dipole-allowed pathway for spontaneous emission: the {$\sigma$-polarized} photon transition $\ket{n \CC} \rightarrow \ket{(n-1) \CC}$. Neglecting the blackbody radiation, the circular state radiative lifetime scales as $n^5$ (typically being on order of $10^{-2}\,\mathrm{s}$ in cryogenic environments)~\cite{2018-Nguyen-PRX}. For an $N$-atom system, the overall system lifetime, limited by spontaneous emission, is therefore on the order of $10^{-2}/N\,\mathrm{s}$. The lifetime should be considered in light of the spin-exchange timescale $1/C^{jk}_{+-}$. Usually, throughout the simulation duration, one would want to simulate at least $10^2$ spin-exchange periods. Assuming that $C^{jk}_{+-}$ are generally on the order of $10^5\,\mathrm{Hz}$ (as is the case in the example system shown in Section~\ref{sec:effective-interactions}), this means that for $N\sim 10$ systems the circular state lifetime is sufficient for at least $10^2$ spin-exchange periods.

The circular state lifetime can be further extended by placing atoms in a small cavity (on the order of a few millimeters) which does not support photon modes with a wavelength corresponding to the spontaneous emission. Theoretical calculations show that such cavities can extend the radiative lifetime of circular atoms to times $\sim 10^1\,\mathrm{s}$~\cite{2018-Nguyen-PRX,2021-Cohen-PRXQ}. A crude ``cavity'', realized by placing the atoms between two capacitor plates, is sufficient to suppress the $\sigma$-polarized spontaneous emission for circular states~\cite{2018-Nguyen-PRX}. This apparatus does not help for elliptical states, because they have additional emission pathways, including $\pi$-polarized photon emissions. However, cavities of more sophisticated shapes can suppress also $\pi$-polarized photon emission, extending the lifetime of elliptical states as well~\cite{2021-Cohen-PRXQ}.

Real experiments also need to contend with other sources of decoherence, such as noise in the external fields, collisions with residual background gas, \emph{etc.}~\cite{2018-Nguyen-PRX}. As an example from practice, in the experiment described in Ref.~\cite{2020-CantatMoltrecht-PRR}, circular $n=50$ Rb atoms were prepared at $10\,\mathrm{\mu K}$ temperature. The measured radiative lifetime was $3.7\times10^{-3}\,\mathrm{s}$ (without an inhibiting cavity); however, additional factors, such as electric field noise and fluctuations of the magnetic field, lowered the overall timescale of irreversible decoherence to $2.7\times10^{-4}\,\mathrm{s}$. Therefore, the magnitude of these factors needs to be calculated separately, depending on the specific experimental approach. 

Another factor is the stimulated blackbody absorption and emission that occurs at nonzero temperatures. While recent experiments have achieved radiative lifetimes of up to $10^{-3}\,\mathrm{s}$ for circular-state atoms at room temperature~\cite{2020-CantatMoltrecht-PRR,2023-Wu-PRL,2024-Holzl-PRX}, operating at cryogenic temperatures remains preferable, to further suppress these effects and extend coherence times.

\subsection{Stability of external fields}

To maintain stable values of interaction coefficients and F\"orster defects over time, precise control of stray electric and magnetic fields is required. For the parameter sets used in our previous examples, we have numerically checked that maintaining the F\"orster defect below $50\,\mathrm{kHz}$ requires stabilizing the electric field to within $10^{-3}\,\mathrm{V/cm}$, and the magnetic field to within $10^{-2}$ Gauss. These accuracy levels are achievable in experimental settings. Electric fields can be measured and compensated with an accuracy of approximately $10^{-5}\,\mathrm{V/cm}$~\cite{1999-Osterwalder-PRL,2014-Huber-NatComm}, while magnetic fields can be controlled with high precision, even down to about $10^{-5}$ Gauss~\cite{2023-Borkowski-RevSciIns,2024-Rogora-PRA}. Hence this level of field stability is experimentally feasible.


\section{Conclusion}\label{sec:conclusion}

In this paper, we have presented a comparative analysis of two methods for simulating spin-1/2 particles using Rydberg states. The first method uses two circular states to represent the spin states, as used in previous proposals, while the second introduces a novel scheme by encoding the spin states in a circular state and an elliptical state. These two approaches result in two distinct ``spin species'' (named ``CC'' and ``CE'' respectively) with different intra- and inter-species spin interactions. We focused on one scheme, where the CE spin is simulated by two states with principal quantum numbers $n',n'+2$; we found that the CC-CC and CC-CE interactions are typically Heisenberg-like, with strong transverse couplings, while CE-CE interactions are more Ising-like, dominated by longitudinal couplings. Combining these two approaches in one setup allows to simulate a variety of two-species spin models, or lattice systems with two sublattices.

We have illustrated the potential applications by providing specific examples of models that could be realized using this framework. This includes: a double lattice system with anisotropic interactions between sublattices, and a pair of coupled Su-Schrieffer-Heeger chains, both of which are relevant to solid-state physics. The setup is experimentally feasible in current ultracold physics laboratories, potentially allowing quantum simulations of larger two- or three-dimensional systems that are difficult to model with traditional numerical methods.

Although our scheme is experimentally viable, certain aspects require further refinement. The inclusion of elliptical states expands the possibilities for spin simulation, but their preparation necessitates adapting existing techniques for generating circular states. Additionally, the current scheme, where the energy differences between pseudo-spin states (${\En_\Ua - \En_\Da}$ and ${\En_\ua - \En_\da}$) are on the order of $10\,\mathrm{GHz}$, is not easily used for qubit applications. For qubit processing, one is typically interested in superpositions of up-spin and down-spin states with specific phase differences. In the presented scheme, the energy difference between the two spin states will be on the order of $10\, \mathrm{GHz}$, leading to phase oscillations with periods around $\approx 10^{-10}\,\mathrm{s}$. This makes it challenging to preserve phase information unless the system can be controlled on nanosecond timescales. However, future developments could mitigate this challenge. For instance, microwave dressing could be used to equate the energies $\En$ of the pseudo-spin states within one or both species (see Ref.~\cite{2018-Nguyen-PRX} for an example). This could be used to enable spin-non-conserving processes or make the system suitable for qubit storage, enhancing its prospects for quantum computing applications.


\begin{acknowledgements}

Jacek Dobrzyniecki thanks Dylan Brown and Giedrius \v{Z}labys from the Okinawa Institute of Science and Technology, as well as Matthias H\"{u}ls from the Forschungszentrum J\"{u}lich, for fruitful discussion. All authors thank Michał
Suchorowski from the University of Warsaw for fruitful discussion.
We gratefully acknowledge the National Science Center, Poland (grant no.~2020/38/E/ST2/00564) for the financial support and Poland's high-performance computing infrastructure PLGrid (HPC Centers: ACK Cyfronet AGH) for providing computer facilities and support (computational grant no.~PLG/2023/016878).
Paula Heim's work was supported by a fellowship of the German Academic Exchange Service (DAAD). Part of this work was supported by the Okinawa Institute of Science and Technology Graduate University.
\end{acknowledgements}


\bibliography{_Biblio}

\end{document}